\documentclass[
 ,twocolumn
 ,secnumarabic
,amssymb, amsmath, aps, nobibnotes,prd]{revtex4}
\usepackage{bm}
\usepackage{graphicx}

\def\centerbmp#1#2#3{\vskip#2\relax\centerline{\hbox to#1{\special
  {bmp:#3 x=#1, y=#2}\hfil}}}

\expandafter\ifx\csname package@font\endcsname\relax\else
 \expandafter\expandafter
 \expandafter\usepackage
 \expandafter\expandafter
 \expandafter{\csname package@font\endcsname}%
\fi
\newcommand{\sig}{\sigma}
\newcommand{\eps}{\epsilon}
\newcommand{\ff}{\phantom{t}}
\newcommand{\ii}{\textit}
\newcommand{\bb}{\textbf}

\newcommand{\be}{\begin{eqnarray}}
\newcommand{\ee}{\end{eqnarray}}
\newcommand{\nn}{\nonumber}
\newcommand{\bit}{\begin{itemize}}
\newcommand{\eit}{\end{itemize}}

\newcommand{\plane}{$(\sigma,\epsilon)$ }
\newcommand{\planes}{$(n_s,r)$ }

\newcommand{\C}{\xi}
\newcommand{\D}{\ff^3\lambda_H}
\newcommand{\LL}{^\ell\lambda_H}
\newcommand{\series}{$\{\epsilon,\sigma,^\ell\lambda_H\}$ }
\newcommand{\M}{m_{\mbox{\scriptsize{Pl}}}}
\newcommand{\sgn}{\mbox{sgn}}
\begin{document}

\title{Dynamics of the Inflationary Flow Equations}
\author{Sirichai Chongchitnan}
\email{sc427@ast.cam.ac.uk}
\author{George Efstathiou}
\email{gpe@ast.cam.ac.uk}

\affiliation{Institute of Astronomy. Madingley Road, Cambridge, CB3 OHA. United Kingdom.}
\date{August 2005}

\begin{abstract}
We present a dynamical analysis of the inflationary flow equations.
Our technique uses the Hubble `jerk' parameter $\xi\equiv (\M^4/ 16\pi^2)
H^{-2}(dH/d\phi d^3H/d\phi^3)$ (where $H$ is the Hubble parameter and
$\phi$ the inflaton) as a discriminant of stability of fixed
points. The results of the analysis are used to explain qualitatively
the distribution of various observable parameters ({\it e.g.} the
tensor-scalar ratio, $r$, and scalar spectral index, $n_s$) seen in numerical
solutions of the flow equations using stochastic initial conditions.
Finally, we give a physical interpretation of the flow in phase-space
in terms of slow-roll motion of the inflaton.
\end{abstract}


\maketitle

\section{Introduction}
Since the first discussions of inflationary cosmology more than twenty years ago, 
a huge number of specific models have been proposed. Most of these
are `phenomenological', though increasing effort has been invested
towards constructing physically motivated models such as brane
inflation (for reviews, see \cite{lr,que}). Given a plethora of
models, one can ask whether aspects of inflation can be studied
without recourse to particular scenarios.

Hoffman and Turner \cite{ht} first addressed this by introducing 
flow equations that govern evolution of observables in any
single-field model of inflation. These flow equations were later
generalised to arbitrary order by Kinney \cite{kin}. In his notation,
the inflationary flow equations are :
\be {d\eps\over dN} &=& \eps(\sig+2\eps)\ff,\nn\\
{d\sig\over dN} &=& -\eps(5\sig+12\eps)+2(\ff^2\lambda_H)\ff,\label{flow}\\
{d\over dN}\ff^\ell\lambda_H &=& \Big[{\ell-1 \over 2}\sig +(\ell-2)\eps\Big]\ff^\ell\lambda_H+ \ff^{\ell+1}\lambda_H\ff.  \ff(\ell\geq2)\nn
\ee
Here the derivative with respect to the number of e-folds, $N$, runs
in the opposite direction to time. The flow equations represent an
infinite dimensional dynamical system of the following parameters
given in terms of the Hubble parameter $H(\phi)$:
\be \eps &\equiv& {\M^2\over 4\pi}\bigg({H'\over H}\bigg)^2,\nn\\\eta &\equiv& {\M^2\over 4\pi}\bigg({H''\over H}\bigg)\ff,\nn\\
\ff^\ell\lambda_H &\equiv& \bigg({\M^2\over 4\pi}\bigg)^\ell{(H')^{\ell-1}\over H^\ell}{d^{\ell+1}H\over d^{\ell+1}\phi}\ff,\label{flow}\\
\sig &\equiv& 2\eta -4\epsilon\ff,\nn
\ee
where $\M$ is the Planck mass and primes denote derivatives with
respect to the inflaton $\phi$. The flow parameters \series are
related to the conventional slow-roll parameters
\cite{lid}
\be \eps_V={\M^2\over16\pi}\Big({V^\prime\over V}\Big)^2\ff,\qquad \eta_V={\M^2\over8\pi}\Big({V^{\prime\prime}\over V}\Big)\ff,\label{sloww}\ee
{\it etc.} defined in terms of derivatives of the inflaton potential
$V(\phi)$.
The validity of the slow-roll approximation is more easily expressed
in terms of these parameters than those of equation (\ref{flow}).  The
relationship between the two sets of parameters is discussed in \S3.

Solving the flow equations gives $\eps(\phi)$ or, equivalently,
$H(\phi)$. These in turn fix the shape of potential $V(\phi)$ via the
Hamilton-Jacobi equation 
\be \big(H'(\phi)\big)^2-{12\pi\over
\M^2}H^2(\phi)=-{32\pi^2\over\M^4}V(\phi)\ff,\label{ham}\ee 
which is equivalent to the Friedmann `acceleration' equation 
\be
{\ddot{a}\over a}={H^2(\phi)}\Big(1-\eps\Big)\ff.\label{acc}\ee 
From
(\ref{acc}), it is clear that inflation is sustained as long as
$\eps<1$. Recently, Liddle \cite{lidd} showed that an analytic
solution to the flow equations, closed at some order, yields a
potential $V(\phi)$ expanded as a rational polynomial.

Using the flow equations, a large number of numerically viable
inflationary models can be readily produced stochastically. However, a
mathematical analysis of the flow equations - an infinite hierarchy of
differential equations - appears daunting. While there have been
discussions of `attractor' solutions \cite{kin, eak} in the past, in
this paper we present a more rigorous analysis of the flow
equations. We begin by exploring the dynamics in the phase-space of
flow parameters \series and classifying all trajectories. In
particular, we are able to explain the complicated motions in the
\plane plane as observed in
\cite{kin,eak}. We then analyse how these motions translate to structures
in the space of observables. Finally, we discuss the physical
implications of our analysis when the Hamilton-Jacobi equation is
incorporated.

\section{Phase-space dynamics}
\subsection{The nature of fixed-points}

Two sets of fixed points of the flow equations are evident from inspection:
\bit
\item \underline{${\eps=0,\ff\sig=}$ constant}- so called `${r=0}$' fixed points, from the fact that the tensor-scalar ratio  $r\propto \eps$ vanishes along that line. 

\item \underline{${\eps>0,\ff\sig=-2\eps, 
\ff^{\ell+1}\lambda_H=\eps(\ff^\ell\lambda_H)}$}- so called 
{`power-law'} fixed point, being a feature of power-law inflation 
from a potential such as $V(\phi)\propto \exp\big(-\sqrt{8\pi}\phi/\M\big)$.
\eit

As a starting point, let us examine the nature of these fixed points
in the simplest two-dimensional projected phase-space.
\subsubsection{Simplest \plane dynamics}

Consider the two lowest order equations in the system
(\ref{flow}). The dynamics in the \plane phase-space clearly depend on
the value of $\ff^2\lambda_H$, the Hubble `jerk' parameter. Following
the notation of \cite{p03}, we shall rewrite it as \be\C &\equiv&
\ff^2\lambda_H \equiv {\M^4\over 16\pi^2}\bigg({H'H'''\over
H^2}\bigg)\ff.\ee Thus we are examining the dynamics of the
two-dimensional system \be {d\eps\over dN} &=&
\eps(\sig+2\eps)\ff,\label{lin1}\\ {d\sig\over dN} &=&
-\eps(5\sig+12\eps)+2\C\ff.\label{lin2}\ee The fixed points in the
\plane system are easy to analyse if $\C$ is treated, for now, as a
constant. By linearising (\ref{lin1}) and (\ref{lin2}), one sees that
the sign of $\C$ discriminates the stability of fixed points in the
\plane plane. The results of a standard fixed point analysis is shown
in Table I.

We draw attention to the `nullclines' where the flow becomes vertical or horizontal:
\be
\mbox{Horizontal nullcline} &:&\quad \sig=-2\eps\ff,\label{hori}\\
\mbox{Vertical nullcline} &:& \quad \eps^2+{5\over12}\sig\eps = {\C\over6}
\ff.\label{vert}\ee
These curves provide useful demarcations of where flows change
direction, and we include them in our phase portraits.

\begin{table*}
\includegraphics[width =17.7cm,  height= 22.74cm]{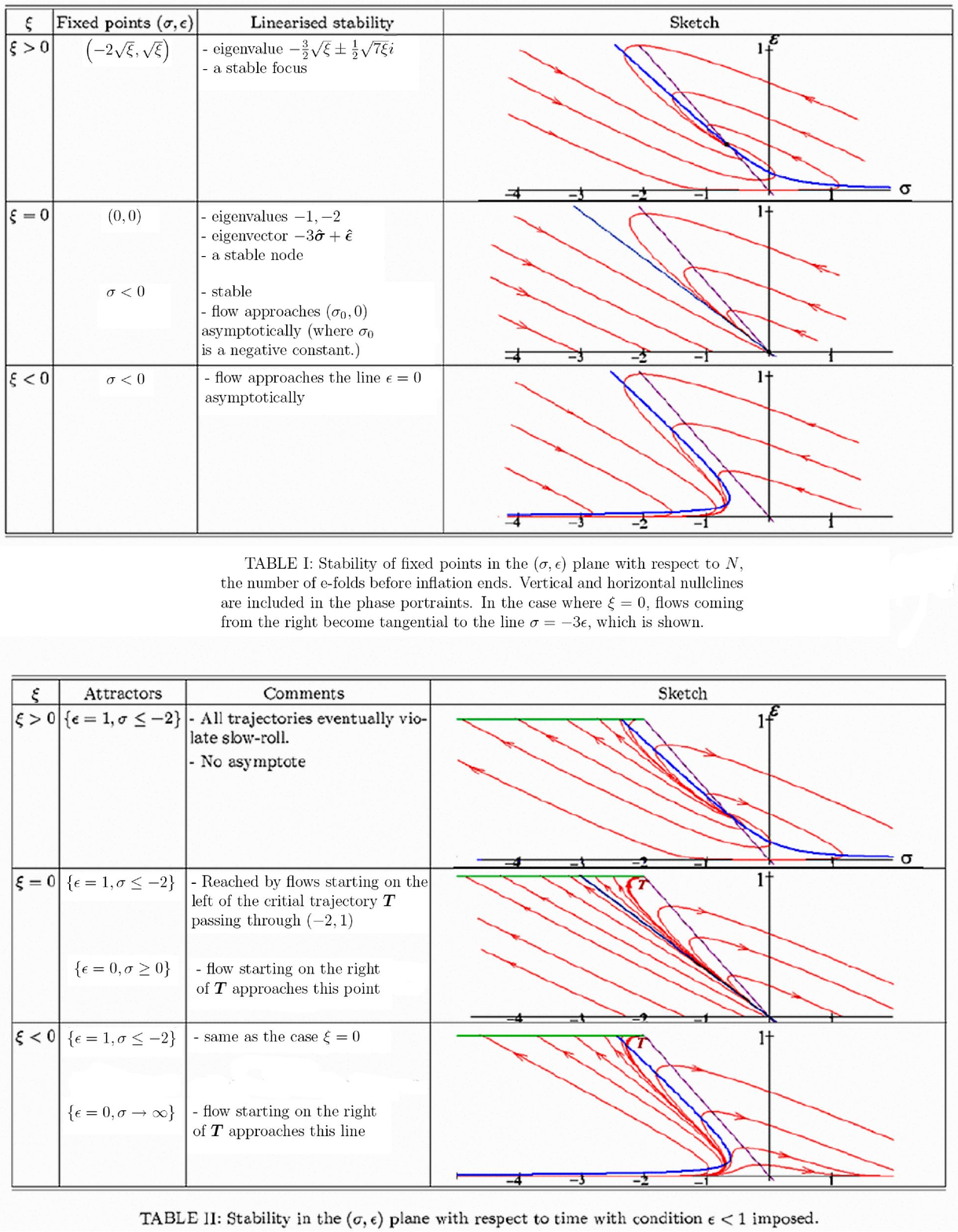}

\end{table*}

\subsubsection{Methodology}

Inflation can be investigated `stochastically' by first selecting a random point in the flow parameter space as an \ii{initial configuration} of a particular universe. Because this is not the same as assigning randomness to \ii{initial conditions} for inflation, we adhere to the phrase `initial configuration' to keep this distinction clear.

An initial configuration is then evolved forward in time (backward in
 e-fold) until inflation ends in one of the following plausible
 ways:
\begin{enumerate} 

\item By achieving $\eps=1$. When this
 happens, we say for convenience that the `slow-roll' condition has
 been violated (keeping in mind that inflation ends because equation
 (\ref{acc}) changes sign rather than because the slow-roll parameters
 $\eps_V$ and $|\eta_V|$ of equation (\ref{sloww}) exceed
 unity). Observables are then calculated a specific number of e-folds
 (which we take to be $55$, unless otherwise stated) before the end of
 inflation by integrating backward in time. This choice is in
 accordance with analyses of references \cite{leach,dod} which
 indicate that for plausible assumptions concerning reheating {\it
 etc}, the observable perturbations were generated 50 to 60 e-folds
 before the end of inflation. Variation of this number is explored in
 \S2.

\item 
By an abrupt termination,  perhaps from intervention of an auxiliary
field in hybrid inflation \cite{lr}, or when open strings become
tachyonic in brane inflation \cite{que}. Because these scenarios
accommodate large number of e-folds during inflation, one identifies
them with an \ii{asymptotic} behaviour of a trajectory. Observables
are calculated along the asymptote. We return to the practical issues
of calculating observables later. 

\end{enumerate}

Because we evolve flows backward in e-fold, the notions of stability
and attraction found in Table I become reversed. This is summarised in
Table II, from which we conclude that given an initial configuration
in \series space, as long as the function:
$$\sgn(\C)=\begin{cases}\ff\ff1\quad \mbox{if} \ff
\C>0\ff,\\\ff\ff0\quad \mbox{if} \ff \C=0\ff,\\-1\quad \mbox{if} \ff
\C<0\ff,\end{cases}$$ remains constant, then the dynamics in the
\series phase-space are completely characterized.

\subsubsection{Dynamics of $\C$}

Having assumed that $\C$ was constant, we can now incorporate aspects
of the dynamics of $\C$ using the function $\sgn(\C)$. If $\sgn(\C)$
changes unpredictably, then the classification in Table II would be
useless. We now show that this is not the case.

Starting from equation (\ref{flow}), one finds that the evolution of
$\C$ has no explicit dependence on $\eps$ \be {d\C\over
dN}=-{1\over2}\sig\C+\D\ff.\label{10}\ee
The solution of equation (\ref{10}) is :
\begin{samepage}
$$\C(N)=\C(N_0)\exp\Big({1\over2}\int_{N_0}^N{\sig(n)}dn\Big)\qquad\qquad$$
\vskip -0.18in
\be\quad\quad+\int_{N_0}^N\D(m)\exp\Big(-{1\over2}\int_{N}^m{\sig(n)}dn\Big)dm\ff.\label{nice}\ee 
\end{samepage}
Now we use equation (\ref{nice}) to deduce the following points:

\bit
\item 
If a trajectory does not violate slow-roll ({\it i.e.} inflation ends
abruptly), then from Table II, there exists some e-fold $N_0$ along
such trajectory such that $\C(N_0)\leq0$ and $\sig(N_0)>0$ for all
$N<N_0$ (since trajectories follow $dN<0$). Assume that typically
$|\!\D|\ll\C$ and using the fact that $\sig(N)$ is unbounded for
asymptotic flow, (\ref{nice}) yields \be \lim_{N\rightarrow-\infty} \C
= 0_-\ff.\label{cool}\ee In other words, if $\C$ starts out negative,
then it is exponentially damped towards zero from below. Slight
deviations may be expected if $\D$ is large, but this is atypical.

\item 
If inflation ends at $\eps=1$ , a similar argument (taking care of the
different integration direction) shows that \be
\lim_{N\rightarrow+\infty} \C = 0_-\quad \ff\ff\mbox{if \ff} \C(N_0)<0
\ff.\label{hot}\ee If $\C(N_0)\geq0$, we also have exponential decay
towards zero,  but only for 55 e-folds. Taking $N_0=0$ to be the end of
inflation, we have :
$$\C(55)=\C(0)\exp\Big({1\over2}\int_{0}^{55}{\sig(n)}dn\Big)\qquad\qquad\qquad\qquad$$
\vskip -0.25in
\be\quad\quad+\int_{0}^{55}\D(m)\exp\Big(-{1\over2}\int_{55}^m{\sig(n)}dn\Big)dm\ff,\label{55}\ee

at 55 e-folds prior to the end of inflation. Unlike the asymptotic
cases, $\C$ cannot have decayed by very much after only 55 e-folds,
and indeed the magnitude of $\C(55)$ can be rather significant
especially when, say, $\D>0$. Typically one finds
$\C(55)\sim\mathcal{O}(10^{-2})$.

\eit

The conclusion is that regardless of its initial value, $\C$ always
exponentially decays towards zero before observables are calculated,
thus typically preserving $\sgn(\C)$. In particular, if inflation ends
with $\C>0$, then the value of $\C$ when observable perturbations were
produced may be significantly non-zero and positive. We return to this point when
we study the structures in the space of observables.

\subsubsection{Numerical method}

We ran a computer code (previously used in \cite{gpe}) that follows
the stochastic prescription to see if we can understand actual
phase-space trajectories. The program selects initial configurations
at random from uniform distributions within the following ranges :
\be \eps_0&\in&[0,0.8]\ff,\nn \\\sig_0&\in&[-0.5,0.5]\ff,\nn \\\C_0&\in&[-0.05,0.05]\ff, \label{window}\\ \LL|_0&\in&[-0.025\times5^{-\ell+3},0.025\times5^{-\ell+3}]\ff, \ff(3\leq\ell\leq10)\nn \\^{11}\lambda_H|_0&=&0\ff,\nn\ee where the
truncation is imposed to close the hierarchy at $\ell=10$. For most
applications of the flow equations, the exact widths of distributions
in (\ref{window}) are unimportant as long as the hierarchy is
convergent. Figure 1 shows \plane trajectories from nine initial
configurations. They are all consistent with the linearised analysis
in Table II.

\begin{figure*}
\includegraphics[width=14.5cm, height=9.5cm]{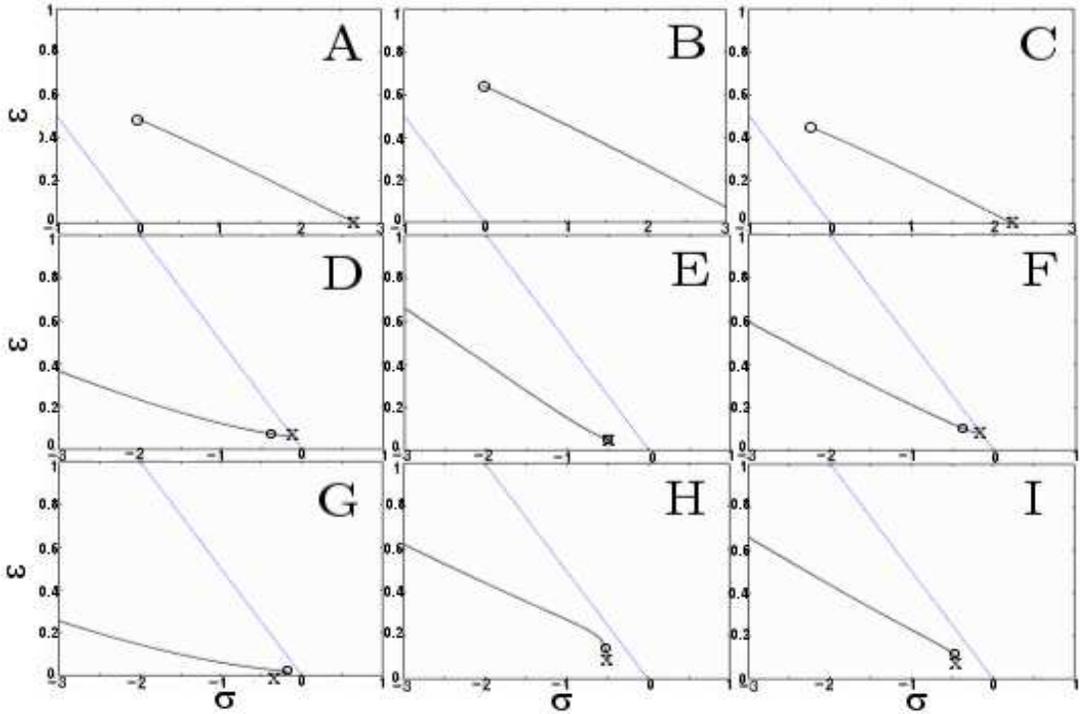}
\caption{Sample of nine \plane trajectories. Circles denote initial random points. Crosses mark where the observable perturbations are produced. In each panel the blue line is the horizontal nullcline (\ref{hori}) Models A,B,C show the most generic behaviour for all values of $\mbox{sgn}(\C)$ (see Table II). The remaining models all violate slow-roll and observables are calculated by backward integration. According to our classification, models D,E and F are clearly consistent with $\C>0$,  whereas $\C<0$ for G, H and I. }\end{figure*}

\begin{figure*}
\includegraphics[width=10cm, height=9cm]{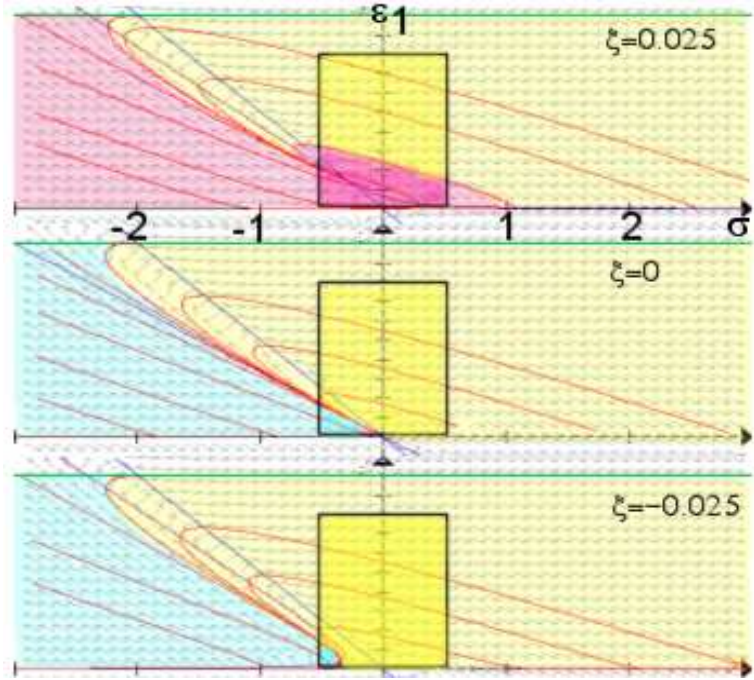}
\caption{Illustrations of critical areas in the \plane plane. 
A random point that starts in a yellow region ends up around
$\eps=0,\sig>0$ (class A in text). Pink and cyan mark potentially
slow-roll violating points whose observables are produced at
$\sig=-2\eps$ (class B) and $\eps=0,\sig<0$ (class C)
respectively. Boxes indicate the initial $(\sig_0,\eps_0)$ window in
equation (\ref{window}). The coloured areas translate to respective
abundance of resulting points from a stochastic experiment as
summarised in equation (\ref{18}). However, this breakdown is
sensitive to the parameter $N_{\mbox{critical}}$ (the number of
e-folds before inflation is terminated abruptly, see text for a more precise
definition). For instance, very large $N_{\mbox{critical}}$ renders
the top panel entirely pink.}
\end{figure*}

By producing 125,000 realisations, we observed the following distribution 
among three classes:
\begin{enumerate}
\item[(A)] Asymptotic flow towards the half-line $\eps=0,\sigma>0$  (for $\approx 90 \%$
of all trajectories).
\item[(B)] Slow-roll violation with observable perturbations produced at the 
power-law line $\sig=-2\eps$ (for $\approx8\%$ of trajectories).
\item[(C)] Slow-roll violation with observable perturbations produced at 
$\eps=0,\sig<0$ (for  $\approx2\%$ of trajectories). 
\end{enumerate} 
This is in agreement with observations in the numerical investigations of
references \cite{kin,lid2,p03}. We have a simple explanation for this distribution.

Figure 2 shows the sketches of Table II with the initial window
(\ref{window}) in the \plane plane superimposed, and regions
colour-coded according to the eventual fate (A, B or C above) of
points starting within them. One deduces that the ratio of these
colour-coded areas determines the final distribution in a given
stochastic experiment. Formally, for the uniform distributions of equation
(16)  we have :
\be \mbox{Percentage of points in class X}\ff\ff\ff\ff\ff\ff\nn\\ \approx \sum_{\sgn(\C)}\frac{\mbox{area for outcome X}}{\mbox{area of }(\sig_0,\eps_0)\mbox{ window}}\ff.\ff\ff\ff\ff\label{18}\ee
where $X = $ A, B or C described above. The sum can be evaluated for
$\C=0,\pm\langle|\C_0|\rangle \approx 0, \pm\max({\C})/2$. Applying
(\ref{18}) to our case where $\pm\langle|\C_0|\rangle=\pm0.025$, one
easily deduces from Figure 2, for example, that the dominant yellow
area in the three panels corresponds to the majority of points
belonging to class (A),  whereas the tiny cyan areas mean that models
belonging to class (C) are very rare indeed.

Figure 2 also shows the effect of changing the size and location of
window $(\sig_0,\eps_0)$ in (\ref{window}). For instance, if
$\max(\eps_0)<0.8$, the percentage of points in case (A) will drop
proportionally.  The fraction of trajectories in each of the
classes (A), (B) and (C) depends primarily on the initial distributions
of the  parameters $\sigma$ and $\epsilon$, rather than the initial 
distributions of  higher-order parameters.

Finally, we comment on a particular aspect of numerical solutions of
the flow equations. Consider the case when $\C_0>0$. From Table II,
one expects
all trajectories to violate slow-roll. In particular, a
trajectory may appear to approach the line $\eps=0,\sig>0$, but
eventually it inevitably intersects the vertical nullcline
(\ref{vert}) and swiftly reverses direction,  thereafter violating
slow-roll. In practice, however, one identifies a flow to
be \ii{asymptotic} (and obeying slow-roll) if it supports inflation
until a prescribed critical number of e-folds is fulfilled. In our
code, this number is set to
$N_{\mbox{\scriptsize{critical}}}=200$. Therefore, trajectories that
could eventually 
violate slow-roll for $N>N_{\mbox{\scriptsize{critical}}}$ are not
identified as such in practice. This explains why in Figure 2 there are
points that will not violate slow-roll even though $\C>0$. Of course,
to select only truly asymptotic flows, one may increase
$N_{\mbox{\scriptsize{critical}}}$ to be, say, 1000 or more. However,
this is at the expense of proportionally increased computation time.

To summarise, the numerical implementation of the flow equations
described here agrees with theoretical analysis of the flow dynamics,
based on the parameter $\C$, described in the previous Section. There
is some dependence of the numerical solutions on  input parameters
such as the e-fold cut-off $N_{\mbox{\scriptsize{critical}}}$ 
and choice of initial \plane window (with weak
sensitivity to distributions of higher order flow variables). These
dependences can be understood theoretically.  In particular, we
interpreted the distribution of numerical outcomes as ratio of areas
in the theoretical phase-space.

\section{Structures in the space of observables}

At about 55 e-folds before the end of inflation, observable
perturbations were produced, setting the values of key inflationary
observables- for example, the scalar spectral index $n_s$ ,
tensor-scalar ratio $r$ and running of spectral index $\frac{dn_s}{d\ln
k}$. To second order, these are related to the inflationary flow
variables by \cite{lpb}:
\be r &\simeq& 16\eps[1-C(\sig+2\eps)]\ff, \label{r}\\ 
n_s&\simeq&1+\sig-(5-3C)\eps^2-{1\over4}(3-5C)\sig\eps+{1\over2}(3-C)\C\ff,\nn\\\label{n}\\
{dn_s\over d\ln k}&=&-{1\over1-\eps}\cdot{dn_s\over dN}\ff,\label{run}\ee 
where $C=0.0814514$. Note that \cite{kin,lid2} define $r$ to be  16
times smaller than in equation (\ref{r}).

We can now make the following classification:
\bit 
\item If inflation ends by  violation of slow-roll conditions, then according to our
earlier analysis, observable perturbations were produced either at (i)
the line $\eps=-2\sig$, or (ii) $\eps=0,\sig<0$. In the first case, we
have an exact power-law inflation, in which case the relation 
\be n_s\simeq1-{2r\over16-r}\ff,\label{a}\ee 
holds \cite{kin}. For case
(ii), one observes \be r\simeq 0\quad,\quad n_s\lesssim1\ff.\label{b}\ee

\item If inflation does not violate slow-roll but ends abruptly, 
then we must have asymptotic flow along $\eps=0,\sig>0$. 
This yields a `blue' spectrum : \be  r\simeq 0\quad,\quad n_s\gtrsim1\ff.\label{c}\ee
\eit

Equations (\ref{a}), (\ref{b}) and (\ref{c}) define three
prominent structures in the \planes plane where we expect points to
cluster strongly. From the analysis of the previous section, we also
know the relative density of these structures: Most points cluster at
(\ref{c}) (class A) , some at the power-law curve (\ref{a}) (class B),
and very few at (\ref{b}) (class C). This behaviour is seen clearly in
Figure 3, which shows a scatter plot from $125,000$ numerical
solutions of the flow equations using the code and initial conditions
described in the previous section.

\begin{figure}

\includegraphics[width=8cm, height=8cm]{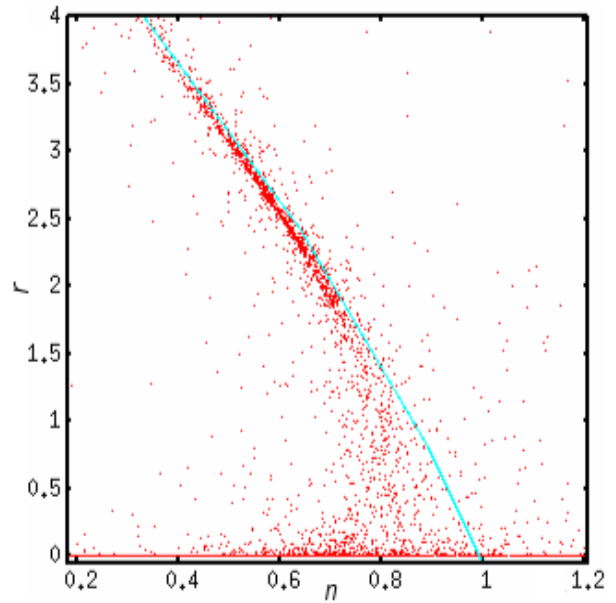}
\caption{Models in the \planes plane generated from the flow equations. 
The line $r=0$ is densely populated. The swathe of points traces the curve 
given by (\ref{a}) (shown by the blue line). The distribution of 
points along this line is discussed in text.}
\end{figure}
 
In terms of the parameter $\C$, we deduce that the swathe
corresponding to equation (\ref{a}) is made up almost entirely of models
in which $\C(55)>0$. The $r=0$ cluster should theoretically be made up
of only those models with $\C(55)\leq0$, although in practice, the
choice of $N_{\mbox{\scriptsize{critical}}}$ means that there are some
points with $\C(55)>0$ on this line as well.

\begin{figure*}
\includegraphics[width=17.4cm, height=7.86cm]{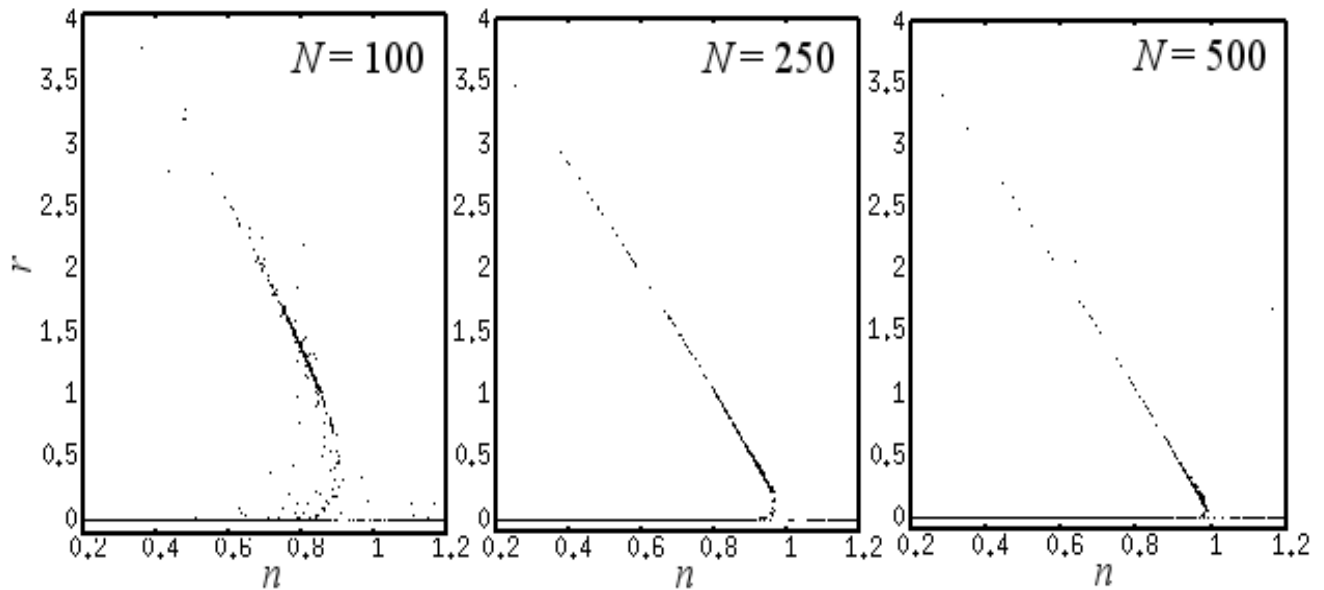}

\caption{Numerical simulations similar to those shown in Figure 3, but
for different values of $N_{\mbox{\scriptsize{obs}}}$. As
$N_{\mbox{\scriptsize{obs}}}$ increases, the swathe of points
around equation (\ref{a}) moves to
lower values of $r$ as described in the text.}
\end{figure*}

The swathe is not evenly populated along the curve (\ref{a}). We have
a simple explanation for this non-uniformity. Starting at the end of
inflation with $\C(0)>0$ and $\eps(0)=1$, the trajectory is traced 55
e-folds backward in time, at which point the trajectory must be near
the line $\sig=-2\eps$ in order to form the swathe in the \planes
plane. From Table I, the $\eps$ coordinate of the power-law fixed
point is $$\eps_*=\sqrt{\C}<1\ff.$$ From (\ref{r}), to first order,
$$r\simeq16\eps(55)\geq16\eps_*=16\sqrt{\C(55)}\ff,$$ so that :   
\be16\sqrt{\C(55)}\ff\lesssim \ff r< 16\ff,\label{bound}\ee
where $\C(55)$ is given by equation (\ref{55}). Only points within these
bounds are expected to lie close to the line defined by equation (\ref{a}).
Intuitively, we should
expect to see some sharp features in the swathe because trajectories
that give rise to the swathe are \ii{suddenly} terminated 55 e-fold by
backward integration in order to calculate $r$. For example, in Figure
3, the lower bound in (\ref{bound}) is found to be 1.7 on
average. For $r \lesssim 1.7$, the points show a large scatter in the \planes 
plane and no longer cluster tightly around the relation defined by 
equation (\ref{a}).

Finally, we comment on the sensitivity of the swathe to the e-fold
limit $N_{\mbox{\scriptsize{obs}}}=55$ when observable perturbations
are generated. As discussed earlier, the investigations of
\cite{leach,dod} favour a range of $N_{\mbox{\scriptsize{obs}}}$
between 50 to $60$, but caution that some models with prolonged
reheating may give $N_{\mbox{\scriptsize{obs}}}$ as high as 100. We
investigate how such an increase affects the
structures in the \planes plane.

An increase in $N_{\mbox{\scriptsize{obs}}}$ only affects slow-roll
violating models.  In the case where observable perturbations are
produced along the line $\eps=0,\sig<0$, extended backward integration
gives rise to a larger spread along the line $n_s<1$ in the observable
plane. On the other hand, for those trajectories approaching the
power-law line $\sig=-2\eps$, an increase in
$N_{\mbox{\scriptsize{obs}}}$ means that $\C$ exponentially decays
towards zero for a longer time, concentrating the points to lower
values of $r$. Furthermore, as $N_{\mbox{\scriptsize{obs}}}$ is
increased, the analogous lower bound to (\ref{bound}) extends to lower
values of $r$. Thus for large values of $N_{\mbox{\scriptsize{obs}}}$,
points cluster tightly around equation (\ref{a}) to values of $r$
close to zero. These effects are evident in Figure 4, which shows analogous
plots to Figure 3, but for numerical simulations with $N_{\mbox{\scriptsize{obs}}}$
is the range $100$ to $500$.

\section{Physical interpretations}

In the slow-roll approximation, the potential shape is parametrized by
its slope and curvature via the {slow-roll} parameters : \be
\eps_V={\M^2\over16\pi}\Big({V^\prime\over V}\Big)^2\quad,\quad
\eta_V={\M^2\over8\pi}\Big({V^{\prime\prime}\over V}\Big)\ff.\ee We
wish to see precisely how the flow parameters \series relate to this
description of the potential shape.

As stated in the introduction, the flow parameters are related to
$V(\phi)$ via the Hamilton-Jacobi equation (\ref{ham}). Dividing it by
$H^2$ gives a useful form : \be \eps-3=-{8\pi\over\M^2}\Big({V\over
H^2}\Big)\ff.\label{v}\ee By differentiating (\ref{ham}), we can
derive two further relations:
\be \pm\sqrt{\eps}(\eta-3)=-2{\sqrt{\pi}\over\M}\Big({V^\prime\over H^2}\Big)\ff,\label{v1}\ee
\be \eta^2-3(\eps+\eta)+\xi=-{V^{\prime\prime}\over H^2}\ff,\label{v2}\ee
where the sign of $\sqrt{\eps}$ matches that of $H^\prime$. Equations (\ref{v}) to (\ref{v2}) relate the amplitude, gradient, and curvature (in that order) of the potential $V$ to the flow parameters.  

Now consider the first slow-roll condition $\eps_V\ll1$. This
condition states that the slope of the potential should be so small
that (i) potential energy dominates kinetic energy of the field
($V\gg\dot{\phi}^2$) and (ii) rolling approaches `terminal' speed so
that $\ddot{\phi}\approx0$. By dividing (\ref{v1}) by (\ref{v}) we
obtain these requirements exactly in terms of flow
parameters:\be\eps_V=\eps\Big({\eta-3\over\eps-3}\Big)^2\ll1\ff.\label{first}\ee

One way to maintain inflation is to ensure that the potential becomes
increasingly flat, so that the inflaton continues approaching the
minimum asymptotically. From (\ref{first}), this can be achieved with
a small $\eps$ regardless of how large $|\eta|$ is. In the \plane
phase-space, this corresponds to those `hybrid' type trajectories
approaching $\eps=0$ regardless of large $\sig$.

Next, the second slow-roll condition $|\eta_V|\ll1$ can be obtained by
differentiating $\eps_V$, and so it places a restriction on the
potential {curvature}.  By dividing (\ref{v2}) by (\ref{v}), we can
recast this condition as \be
|\eta_V|=\bigg|{3(\eps+\eta)-{\eta^2}-\C\over
3-\eps}\bigg|\ll1\ff.\label{second}\ee(Note the role of parameter $\C$
as a constraint on the curvature.) In particular, we see that a
universe with sufficiently large negative $\eta$ will violate
(\ref{second}) even though the first slow-roll condition (\ref{first})
is satisfied. This shows why potentials with sufficiently large
negative curvature (such as those associated with  spontaneous symmetry
breaking or  new inflation \cite{as,linde}) are doomed to violate
slow-roll. As the inflaton gains speed along a concave-down potential,
its `image' in the phase-space is a flow towards the line
$\{\eps=1,\sig\leq-2\}$ as shown in Table II.

In summary, one can understand the correspondence between inflaton
motion along a potential and the flow in \series phase-space simply by
using the flow parameters to characterize the gradient and curvature
of the potential.

\section{conclusions}

The inflationary flow equations have been used by a large number of
authors to draw inferences about observable parameters. When solving
the flow equations, one finds favoured regions of observable
parameters due to dynamical attractors. The purpose of this paper has
been to examine the dynamics of the flow equations analytically. In
particular, we have shown that the Hubble `jerk' parameter $\C$
effectively discriminates the stability of fixed points in the flow
parameter phase-space, while the dynamics of $\C$ itself can be 
described simply in terms of exponential decay. Using this technique, we were
able to explain complicated motions in the \series phase-space
analytically. The distribution of points and stability of structures
in the plane of observables can also be understood in terms of
dynamics of $\C$. Comparing our analytic work with numerical solutions
of the flow equation, we found that they were in good
agreement. Finally, we have given exact relations showing how the
trajectories in the phase-space of flow variables correspond
physically to motion of the inflaton along a potential.

The main features in the plane of observables are generally
insensitive to the initial configuration (i.e. choice of values
$^\ell\lambda_H\big|_0$) - although there is a dependence on the
parameter $N_{\mbox{\scriptsize{critical}}}$ which can be understood
dynamically. Thus we believe that we understand the main features of
the solutions of the flow equations. Nevertheless, to interpret
diagrams such as figures 3 and 4 in any statistical sense requires an
understanding of the thorny issue of measures and how initial
conditions for inflation might translate into distributions of initial
configurations. We have not attempted to address these problems 
in this paper. For some speculations, see references
\cite{vil,tegmark}.

\vskip 0.1 truein

\noindent
{\bf Acknowledgements:}
SC acknowledges the Dorothy Hodgkin scholarship from PPARC. This work
has been supported by grants from PPARC.

\end{document}